\documentclass{nature}
\usepackage{bm}
\usepackage{graphicx}

\title{Quasi-droplet Microbubbles for High Resolution Sensing Applications}

\author{Yong Yang$^1$, Jonathan Ward$^1$ and S\'{i}le Nic Chormaic$^1$}

\begin{document}
\maketitle
\begin{affiliations}
\item Light-Matter Interactions Unit, OIST Graduate University,\\ Okinawa, 904-0495, Japan
\end{affiliations}

\begin{abstract}
Optical properties and sensing capabilities of fused silica microbubbles were studied numerically using a finite element method. Mode characteristics, such as quality factor (Q) and effective refractive index, were determined for different bubble diameters and shell thicknesses. For sensing applications with whispering gallery modes (WGMs), thinner shells yield improved sensitivity. However, the Q-factor decreases with reduced thickness and this limits the final resolution. Three types of sensing applications with microbubbles, based on their optimized geometrical parameters, were studied. Herein the so-called quasi-droplet regime is defined and discussed. It is shown that best resolution can be achieved when microbubbles act as quasi-droplets, even for water-filled cavities at the telecommunications C-band.
\end{abstract}


\section{Introduction}
The benefits of high quality (Q) factors in whispering gallery mode resonators (WGRs) have been studied intensively during the last several decades \cite{Vahala2003}. This unique feature of WGRs is a key factor in the study of low-threshold microlasers \cite{He2013}, for nonlinear effects \cite{Spillane2002,Wu2010}, in cavity quantum electrodynamics \cite{Aoki2006,Park2006}, and for optomechanics \cite{Kippenberg2007}. WGRs are typically micron-scale dielectric structures that can confine light internally by a process of continuous total internal reflection. Light circulates around the boundary and forms whispering gallery modes. Wavelengths of WGMs are highly dependent on the geometrical size and refractive index of the resonator.  Resonant frequencies of WGMs are also very sensitive to external influences and this leads to ultrahigh sensitivity in sensing applications. By now, sensing has been accomplished using various WGRs, such as microspheres \cite{Arnold2003} and microtoroids \cite{Vahala2003,Armani2007}. A large diversity of physical quantities can be sensed, e.g. biochemical changes \cite{Arnold2003,Vollmer2008}, gas \cite{Gregor2010}, temperature \cite{Dong2009,JonathanWard2013,Henze2013}, pressure \cite{Henze2011}, and force \cite{Ilchenko1998}. Due to the high Q and small mode volume of WGRs, even single molecule detection has been achieved \cite{Armani2007,Vollmer2008}.\\
For the sake of general discussion, let us  consider refractive index sensing as an example. The sensitivity of a resonator relies on the portion of electromagnetic (EM) field distributed  outside it, or in another words, the tunneling depth of the evanescent field. To extend the evanescent field, several methods have been developed, such as PDMS-coated microspheres \cite{Dong2009} or plasmonic enhancement in metal-coated microresonators \cite{Xiao2010,Disfani2012,Xiao2012}. By having different dielectric layers, WGM EM field distribution can be tailored. Liquid core optical ring resonator (LCORR) sensors \cite{White2006a} are an alternative type of WGM resonator. Such devices can  be viewed as hybrid microresonators,  since the evanescent light field can penetrate into the liquid in the core. In LCORRs, however, a high Q is maintained because most of the WGM energy still propagates in the shell structure. Therefore, LCORRs have outstanding sensing properties.\\
In this paper, one type of LCORR is discussed, namely the microbubble resonator \cite{Sumetsky2010,Watkins2011}. They are made by locally heating a fused silica microcapillary with a CO$_2$ laser while internally pressurizing the cavity. By controlling the size of the CO$_2$ laser heating zone, a spherical shape, with a controllable shell thickness, can be created. Single- and double-pass structures, i.e. spherical shells with one or two openings, can be made. For double-pass structures, liquid can be injected through the cavities using a syringe pump. Similar to other LCORRs, the core liquid is sensed by an EM field traveling internal to the cavity core. Sensitivity of the WGM to changes in the liquid core can be improved by making the shell thinner, thereby increasing WGM EM field intensity in the liquid. However, when the shell thickness decreases to near or less than the wavelength of the light propagating in the WGM, tunneling can occur, causing the resonant line width to broaden, thereby limiting the total resolution. As a result, a tradeoff must be achieved between shell thickness and the fixed size of the microbubble in order to optimize the Q-factors and, hence, the sensitivity of the device. To our knowledge, this optimization of the LCORR has not been reported in the literature to date. Herein, the axi-symmetrical finite element model (FEM) is used to investigate the optical mode properties of microbubble WGRs. Propagation constants of different order modes are strongly related to bubble size and shell thickness. Therefore, controllability of the  modes is made possible by changing the coupling conditions. When the shell thickness is subwavelength and the bubble contains a high-index liquid inside, a quasi-droplet regime is defined.  Based on different physical sensing applications, optimized parameters for microbubbles are determined.\\
\section{FEM simulations of microbubbles}
\begin{figure}
\centerline{\includegraphics[width=8cm]{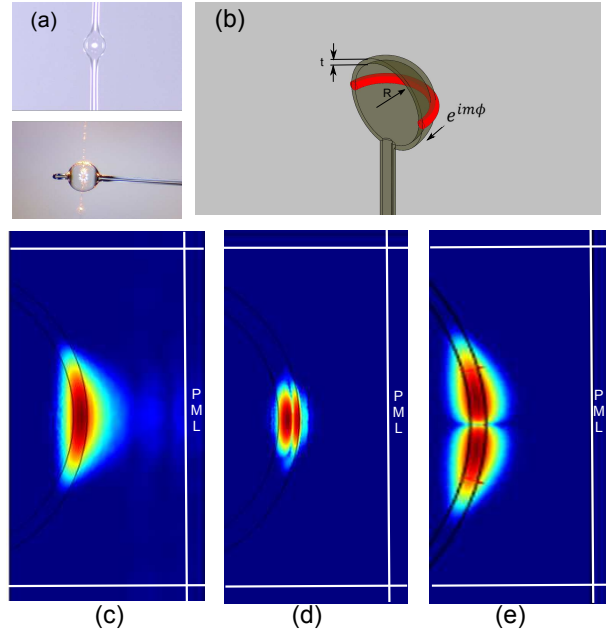}}
\caption{(a) Images of double-pass and single-pass microbubbles. (b) Whispering gallery modes propagate along the surface of a microbubble (red trace). For purpose of illustration, a microbubble is cut transversely along the polar axis. $R$ is the outer radius of the microbubble and $t$ is the shell thickness. The mode number, $m$, determines the relationship between the azimuthal field, $\mathbf{E}$, and the azimuthal coordinate, $\phi$, as $\mathbf{E} \propto \exp(im \phi)$. Therefore, the solution of the 3D FEM problem can be reduced to 2D along the bubble's symmetry axis. Radial mode distribution patterns are shown in (c), (d) and (e). (c) is the first radial order fundamental mode,  (d) is the second radial order mode, and (e) is a higher transverse mode. All are quasi-TE modes. To derive Q-factor values, perfectly matched layers (PMLs) are required (c) -- (e). }
\label{fig:scheme}
\end{figure}

\begin{figure}
\centerline{\includegraphics[width=8cm]{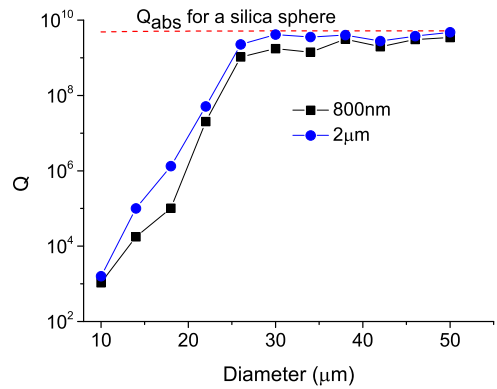}}
\caption{Q-factors of microbubbles drop exponentially with decreasing radii due to greater radiation loss for smaller WGM cavities. This is illustrated on the plot with quasi TE fundamental modes. Two different shell thicknesses, 800 nm (black square) and 2 $\mu$m (blue circle), are compared. The red dotted line represents the absorption Q limit of a silica microsphere calculated using the same FEM simulation method with a diameter of 50 $\mu$m. Due to material loss, the Q-factor is limited by the absorption Q.}
\label{fig:qvr}
\end{figure}

\begin{figure}
\centerline{\includegraphics[width=8cm]{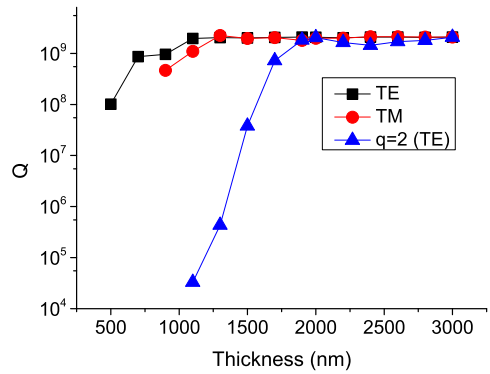}}
\caption{Three different types of modes coexist in microbubbles, represented as squares (TE), circles (TM) and triangles ($q=2$ TE). However, due to inner surface tunneling loss, higher radial modes have lower Q-factors than lower modes. The diameter of the microbubble for this plot is 50 $\mu$m. The TE mode is higher than the TM mode, especially when the shell is thin. The maximum Q-factor is limited by the silica absorption. }
\label{fig:qvt}
\end{figure}

\begin{figure}
\centerline{\includegraphics[width=8cm]{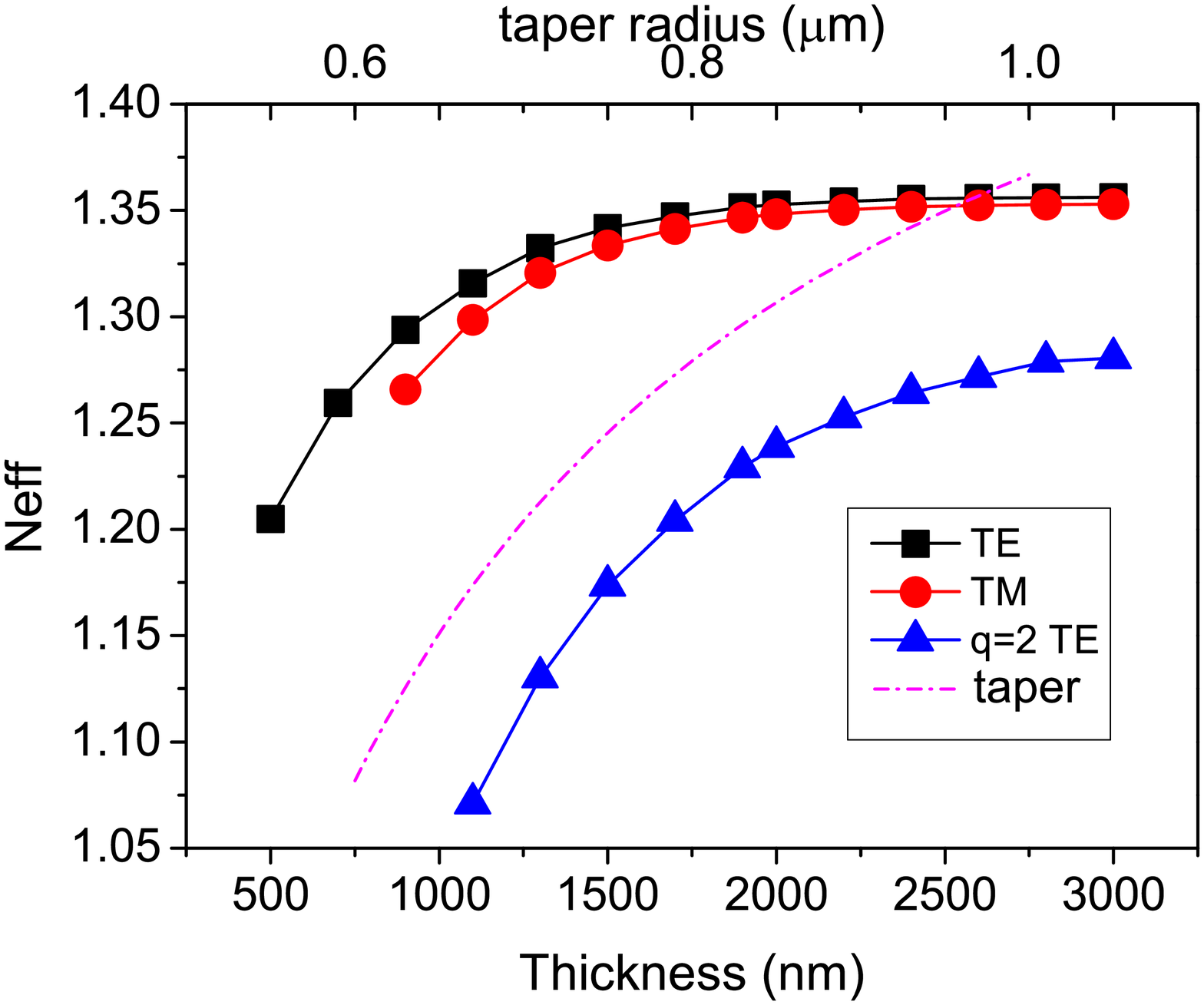}}
\caption{Effective index of a 50 $\mu$m diameter microbubble for different shell thickness and different modes.  Black squares are the fundamental TE mode and red circles are the TM mode. The effective index of the second radial order is plotted in blue triangles for a shell thickness from 1.1 $\mu$m, where the air-filled bubble starts to support high order modes. The taper effective index for a fiber waist of 0.5-1.0 $\mu$m radius is also presented (dashed pink line). Once the geometry of a microbubble is set, a proper taper size can be chosen to satisfy the phase matching condition.}
\label{fig:taper}
\end{figure}

 Images of typical microbubbles are shown in Fig. \ref{fig:scheme}(a) and the schematic cross-section of a microbubble is shown in Fig. \ref{fig:scheme}(b).  For simplicity we assume that the microbubble is a spherical shell formed by fused silica and surrounded by air with  core materials that can be varied. A working wavelength of 1.55 $\mu$m was chosen, as it is commonly used in WGM experiments. FEM simulations of a 3D structure consume a lot of computational resources even for micron-scale objects. The microbubble is rotationally and axially symmetric, so by utilizing a newly developed FEM \cite{Oxborrow2007}, the 3D problem is reduced to 2D and is solvable in seconds with smaller computational memory requirement.
The method in \cite{Oxborrow2007} is based on the weak form of the Helmholtz equation \cite{Cheema2013},  given by:
\begin{equation}
\int dV ((\nabla \times \tilde{\mathbf{H}} ^{\ast})\epsilon ^{-1}(\nabla \times \mathbf{H})-\alpha(\nabla \cdot \tilde{\mathbf{H}} ^{\ast})(\nabla \cdot \mathbf{H})
+c^{-2}\tilde{\mathbf{H}}\cdot\frac{\partial ^{2}H}{\partial t^{2}})=0
\label{eq:weakform}
\end{equation}
where $\epsilon$ is the effective permittivity and $\alpha$ is the penalty factor first introduced in \cite{Oxborrow2007}.\\
In  spherical coordinates $(r, \theta, \phi)$, WGMs propagate azimuthally to the rotational symmetric axis, as illustrated in Fig. \ref{fig:scheme}(b).  This gives rise to a field phase varying term, $\exp(im\phi)$. Here, $m$ is the azimuthal mode order of the WGM. In the simulation, $m$ is varied and eigenfrequencies of corresponding fundamental modes are determined for different EM field distributions along the radial direction (Fig. \ref{fig:scheme}(c)--(e)). The effective index of the mode is estimated by $N_{eff}=m\lambda /2\pi R$, where $R$ is the outer radius of the microbubble.\\
For WGRs, the Q-factor is a very important parameter. The total intrinsic loss of a WGR originates from radiation loss (tunneling loss), material loss, surface roughness, and contamination. Here, only radiation and material losses are considered. The surface roughness is very small due to the fabrication method used. In experiments, a high Q absorption limit for microbubbles has been reported \cite{Bahl2013}. Radiation loss is caused by leakage from evanescent light into a free space mode. The upper and lower bounds of the Q-factor can be estimated with a closed resonator model. In this work, a more precise method was used, in which a perfectly matched layer (PML) along the boundary of the computation domain is introduced, see Fig. \ref{fig:scheme}(c) -- (e). A properly set PML can be treated as an anisotropic absorber, simulating radiation tunneling to infinity within a limited domain calculation space. An accurate determination of the Q-factor in a microsphere has been reported recently using this modified method \cite{Cheema2013}. In order to match the model with a realistic situation, material absorption is introduced as an additional imaginary part to the resonator permittivity. For fused silica and a 1.55 $\mu$m wavelength, the imaginary part is estimated to be $\epsilon_{i}=-3.56\times 10^{-10}$, which is calculated from the absorption coefficient. As will be demonstrated in the following, the radiation loss is dominant when the diameter of the microbubble is less than $\sim 30 \mu$m. The Q-factor exponentially increases with diameter such that it is saturated when $R>$30 $\mu$m, and is then only limited by the material absorption loss.\\
When solving the eigen-equations using FEM software, such as COMSOL\copyright{ },  with complex material permittivity and PMLs, the eigenfrequencies $(f_r)$ are complex, with the real parts representing resonant frequencies and the imaginary parts representing total intrinsic losses. Therefore, the Q-factor is defined as $Q=Re(f_r)/2Im(f_r)$. For the material absorption term, the upper bound is limited to around $10^{9}$, which will be shown in the following simulation results. For investigating WGM properties, air ($\epsilon=1$) is initially chosen as the core material in this section.\\
Microbubbles with different diameters (10-60 $\mu$m) and wall thicknesses (500 nm to 3 $\mu$m) were simulated(Fig. \ref{fig:qvr}). Similar to solid microspheres, Q-factors increase with diameter. Exponential curves for diameters below 30 $\mu$m indicate that radiation losses dominate. When microbubbles are larger, radiation losses diminish and become negligible compared to material losses. As expected, Q-factors for large diameters do not exceed absorption Q-factors of solid microspheres ($10^{10}$).\\
 Two microbubbles with different shell thicknesses were compared. It is clear that when the shell is thinner, the mode tunnels more into the core, increasing the radiation loss and reducing the Q-factor. Therefore, to design high Q-factor microbubbles, larger diameters and thicker shells are required. In the following calculations, 50 $\mu$m has been selected as a reasonable microbubble size, since it can be easily fabricated and shell thickness can be controlled in fabrication \cite{JonathanWard2013}.\\
Before discussing the shell thickness relationship, it is necessary to note that in addition to fundamental modes, other modes also exist in microbubbles (c.f. Fig.\ref{fig:scheme}(d) and (e)). These can be denoted as higher radial modes ($q=2,3,\ldots$) or higher azimuthal modes ($l=m\pm1,m\pm2,\ldots$). For the first radial, fundamental TE mode, when the thickness is less than 1 $\mu$m, the Q-factor drops extremely sharply (Fig. \ref{fig:qvt}). The TM mode has a lower Q-factor and it drops when the shell thickness is less than 1.3 $\mu$m. At a wall thickness of $\sim$ 600 nm, the Q of the TM mode drops to a very low value, implying that the microbubble can only hold the TE mode. For shell thickness less than 500 nm , even the TE mode has a very low Q-factor and microbubbles cannot hold any high Q WGMs. It is worth noting that, when the thickness is larger than the working wavelength (1.55 $\mu$m in this paper), microbubbles can even hold second order radial modes ($q=2$). Radial mode distribution is dependent on the medium along the radial direction. If the shell becomes even thicker, microbubbles should be able to hold even higher radial modes until they become the same as solid microspheres. In other words, single radial mode operation is only possible for microbubbles with subwavelength shells.\\
For real sensing applications, light has to be coupled in and out of the WGR for detection. Many coupling methods have been developed and, among them, tapered fibers exhibit high efficiency as evanescent probes that are widely used for WGRs \cite{Cai2000}. In order to effectively couple light, the cavity mode must be sufficiently spatially overlapped with the mode from the tapered fiber and a phase matching condition must be met, i.e. the effective index of the WGM must equal that of the tapered fiber mode. To verify efficient coupling in the microbubble tapered fiber system, the effective index of a 50 $\mu$m microbubble fundamental TE mode was calculated (Fig. \ref{fig:taper}). For comparison, the index for different tapered fiber diameters is also shown. Note that this index is calculated when the fiber is in contact with the microbubble. To tune the effective index, one can control the taper/microbubble gap or change the taper diameter.\\
From Fig. \ref{fig:taper} it is clear that for thinner microbubbles, the effective index decreases, which is also due to more EM field distributed in the core. For a 50 $\mu$m microbubble, the effective index of the TE mode varies from 1.20 to 1.35. Phase matching can be realized if the taper diameter is controlled between 1.4 $\mu$m and 1.8 $\mu$m. The second order mode has an even lower effective index, ranging from 1.05 to 1.28, so a thinner taper is required to efficiently couple with this mode. In the following discussion, we assume that only the first order fundamental TE mode is of concern, since it has a larger Q-factor than the higher order modes. Efficient coupling to such modes is realized and controlled by selecting the size of tapered fiber.\\
\section{Quasi droplet regime of microbubbles}
\label{sec:droplet}
\begin{figure}
\centerline{\includegraphics[width=8cm]{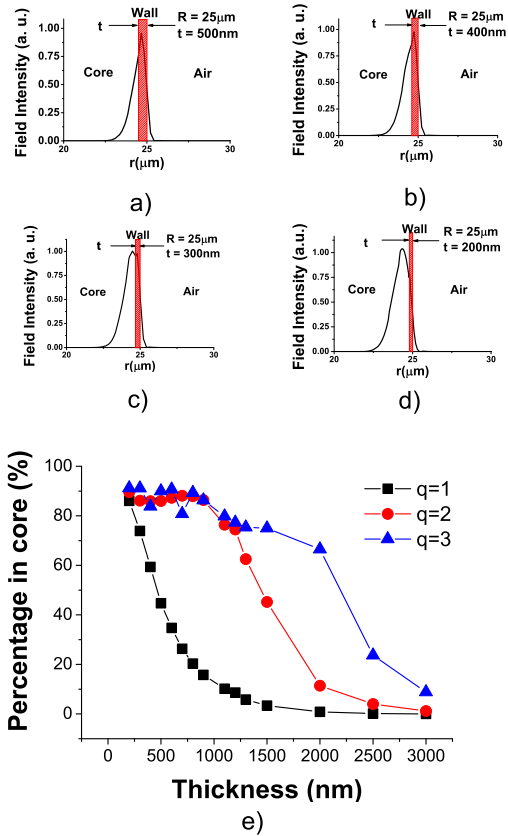}}
\caption{Radial field distribution for  a water-filled, 50 $\mu$m microbubble. From (a) to (d) shell thickness decreases from 500 nm to 200 nm. The $y$-axis represents  $|E|^2$ along the radius $r$, for the TE fundamental modes. When the shell is less than 300 nm thick, the maximum shifts completely inside the core and this is defined as the quasi-droplet regime. In (e), the percentage of light intensities for different radial modes inside the core are calculated. It can been seen that higher order modes have more light distributed in the core, even for microbubbles with thicker shells.}
\label{fig:droplet}
\end{figure}

\begin{figure}
\centerline{\includegraphics[width=8cm]{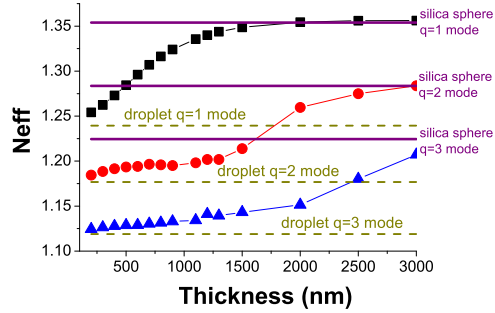}}
\caption{The effective refractive indices of microbubbles with shell thicknesses from 200 nm to 3 $\mu$m. First (black squares), second (red circles) and third (blue triangles) order radial modes are shown and compared with those of a liquid droplet (horizontal dashed lines) and a silica microsphere (horizontal solid lines) of the same diameter. Water was chosen as the liquid and the structures are 50 $\mu$m in diameter.}
\label{fig:droplet2}
\end{figure}

The foregoing discussion has been centered on WGM properties of empty (air-filled) microbubbles; however, it is of more significance to investigate the microbubbles filled with liquid. Since the refractive index of liquid is higher than air and a spherical boundary can be shaped if the liquid forms a droplet, WGMs can be found in such a droplet. Such droplet WGRs have been studied for lasing \cite{QIAN1986} and nonlinear effects \cite{Uetake2002}. Indeed, droplet-like WGMs can also be found in microbubbles.\\
 If the shell of a microbubble is very thick, most of the EM field of the mode will propagate within the shell, so the microbubble behaves like a solid microsphere. As it gets thinner, mode gets extended more into the core. For an extreme situation, $t\rightarrow0$, the mode is almost entirely propagating in the liquid core, where a droplet like condition is satisfied, given that the inner boundary is spherical. Between these two situations, there exists a region where the shell starts to lose the ability to confine WGMs. This region has been dubbed the quasi-droplet regime \cite{Lee2011}. It is worth noting that higher radial modes occupy more space than first order modes; therefore, they cannot exist in air-filled microbubbles with very thin shells. However, the core of a liquid-filled bubble provides the space required for higher order modes to propagate, so higher modes can be supported in thin-walled, liquid-filled bubbles. For example, in Lee \emph{et al.} \cite{Lee2011}, the quasi-droplet regime mentioned is only for the $q=2$ mode. Therefore, it is very important to clarify the definition of the quasi-droplet regime for different modes. In the following discussion, the quasi-droplet regime is defined in terms of the effective index.\\
To numerically simulate this regime, the permittivity of the inner core material was replaced with a liquid (for example,  water with $\epsilon_{real}=1.33^{2}=1.7698$) and substituted into the FEM equation (Eq. \ref{eq:weakform}). The absorption of the liquid can be omitted in the model for calculating mode eigenfrequency and field distribution, since it only changes the imaginary part of the eigenfrequency simulated by FEM software. This will be discussed in the next section. The radial EM field distribution for  $q=1$ was determined by extracting the intensity value along the radial direction (c. f. Fig. \ref{fig:droplet}(a)-(d)). It is distributed along the radius, with part inside the core area, part in the shell and an evanescent tail tunneling to the outer environment.\\
The estimated percentage of the WGM's EM field in the core was found by integrating the EM field intensity in the core and shell separately. The percentage of  energy in the core for the first three radial modes for  a shell thickness varying from 300 nm to 3 $\mu$m is calculated for fixed diameter microbubbles, see Fig. \ref{fig:droplet}(e). It can be seen that, when the shell thickness is 500 nm or less, up to 85\% of the $q=1$ WGM propagates in the core, while if the shell thickness is more than half of the working wavelength (1.55 $\mu$m), more than 80\% of the light travels in the shell. \\
To describe the quasi-droplet regime more precisely, we resorted to a quantified definition for the fundamental TE mode. The idea is based on the well-known interpretation of the radial mode number. For a liquid microsphere, i.e. a droplet, the radial field distribution has a maximum inside the droplet close to the boundary. Analogous to a droplet, when the peak is inside the core of a microbubble, the core is equivalent to the droplet while the shell is the new boundary. This can be used as a criterion for the quasi-droplet regime. It can be physically interpreted that light is traveling in the water and the field distributed in the shell is the evanescent component tunneling into the shell. According to this definition, for a shell thickness of less than 300 nm, the microbubble is driven into the quasi-droplet regime for its fundamental TE mode. However, such defined thickness does not apply for the $q=2$ and $q=3$ modes, as those modes have multiple peaks along the radial direction so they are more complicated than the $q=1$ case. Where the percentage of the EM field for the $q=2$ and $q=3$ modes are plotted, even when the shell is as thick as 1 $\mu$m and 1.5 $\mu$m, respectively, the proportion of the EM field in the core does not drop to less than 80\% (Fig. \ref{fig:droplet}(e)). \\
To have a general definition that applies to different radial modes, the effective indices for the $q=1, 2$, and $3$ modes in microbubbles were calculated. For comparison, the effective index of a droplet of the same diameter was simulated and is shown together with those for a solid silica microsphere (Fig. \ref{fig:droplet2}). The effective index for a microbubble in the quasi-droplet regime, as defined above, is only slightly higher than for the droplet modes, proving that, in this case, the shell is negligible and the microbubble acts like a droplet. For thicker shells $>$2.5 $\mu$m, the effective indices of all bubble modes are the same as the corresponding modes in a solid silica microsphere of the same size. The $q=1$ mode does not reach the droplet index unless the shell thickness is less than 500 nm. On the other hand, higher order modes, especially the $q=2$ modes, exhibit a much wider range of effective indices corresponding to those of the droplet. The effective index of the $q=2$ mode changes abruptly to more closely resemble a solid microsphere when the shell is thicker than 1.5 $\mu$m. Accordingly, a new definition for the quasi-droplet regime could be the range of shell thicknesses until the point where the effective index starts to rapidly approach that of a silica microsphere.\\
So far the quasi-droplet has been described in two ways and we have shown that, for a very thin shell, a microbubble filled with liquid behaves very similarly to a droplet WGR.  This may be very interesting in applications such as sensing or nonlinearity. The quasi-droplet resonator has advantages since its shape is protected by the shell. Changes to resonator shape through surface evaporation can thereby be avoided, and coupling to external waveguides is easier instead of low efficiency, free space excitation \cite{QIAN1986}.\\
\section{Optimizing microbubble geometry for high resolution sensing applications}
Sensing applications for WGRs are mostly based on the following principle: changes in the environment cause a shift in the resonant frequency of the whispering gallery modes. This is detected by scanning and recording the transmission spectrum of the WGR. Sensitivity, $S$, is defined as the shift rate of WGMs. However, if the line width of the resonance dip is broad, the shift cannot be resolved, thereby limiting the resolution of WGR-based sensors. Unfortunately, improving sensitivity is nearly always in conflict with achieving higher Q-factors. In order to overcome this limitation, it is important to have a method of determining optimal parameters for WGR sensors. In general, the resolution, $\Re$, is defined as the following:
\begin{equation}
\Re=\frac{\lambda}{Q}\left(\frac{\partial{\lambda (U)}}{\partial{U}}\right)^{-1}
\label{eq:res}
\end{equation}
where $U$ is the physical quantity (e. g. temperature or pressure) that causes the frequency shift, and $\lambda$ is the working wavelength. $\partial \lambda(U) /\partial{U}$ is the frequency shift caused by the change of the physical quantity, i.e. the sensitivity. In practice, there are two ways to induce the frequency shift and both will be addressed in the following  subsections.\\
\subsection{Pressure sensing}
\label{sec:pressure}
\begin{figure}
\centerline{\includegraphics[width=8cm]{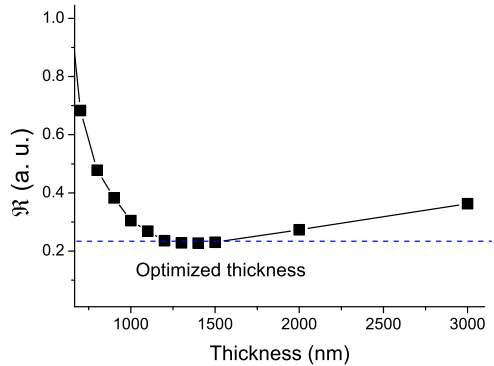}}
\caption{Resolution $\Re$ versus shell thickness for a microbubble used in pressure sensing. The diameter of the microbubble is 50 $\mu$m. The blue line shows the minimum range of resolution. It corresponds to an optimized thickness around 1.4 $\mu$m.}
\label{fig:pressure}
\end{figure}

The resonant frequency of a microbubble WGR can be tuned by manipulating the compression or tension of the device \cite{Sumetsky2010a}. Alternatively, a mode frequency shift of hundreds of GHz can be generated in a single-pass microbubble by gas pressure \cite{Suter2007}. Two different mechanisms are dominant. The first is size expansion by applying aerostatic pressure from inside the bubble. The second is a possible refractive index change due to strain and stress on the resonator material. For a given material the elasto-optic coefficient ($C$) and shear modulus ($G$) are constants, so the pressure sensitivity is given by \cite{Henze2011} (neglecting external pressure):
\begin{equation}
\frac{d\lambda(p_i)}{\lambda}=\frac{2n_0(R-t)^3+12CG(R-t)^3}{4Gn_0(R^3-(R-t)^3)}p_i,\\
\label{eq:pressuresense}
\end{equation}
where  $n_0$ is the refractive index. From Eq. \ref{eq:pressuresense}, the sensitivity of pressure sensing is proportional to the geometrical parameters $R$ and $t$ and a relative sensitivity, $S_r$, is defined by:
\begin{equation}
S_r=\frac{\partial \lambda}{\partial p}\propto \frac{R^3}{R^3-(R-t)^3}.
\label{eq:pressure}
\end{equation}
For simplicity, in the following content, $S_r$ is used for sensitivity, and relative sensitivity is not distinguished from absolute sensitivity. Using Eq. \ref{eq:res} and Eq. \ref{eq:pressure}, $\Re$ can be calculated. Note that since it is deduced from relative sensitivity, it is a relative resolution. Again, for simplicity, $\Re$ represents relative resolution in the following sections. It was plotted as a function of shell thickness, incorporating the Q-factor plotted in Fig. \ref{fig:qvt} and this is shown in Fig. \ref{fig:pressure}. It can be seen that the best resolution is obtained with a shell thickness of about 1.4 $\mu$m. The resolution worsens when the shell thickness is less than 1 $\mu$m. This is due to the exponentially decreasing value of Q with decreasing shell thickness. When the shell is thicker than 1.5 $\mu$m, the resolution also worsens, as the sensitivity to pressure diminishes with increasing shell thickness. The Q-factor reaches the material limit when the shell thickness is more than 1.4 $\mu$m. Sensitivity changes as the inverse cube of the shell thickness, which causes less severe deterioration of the resolution.\\
\subsection{Refractive index sensing}
\label{sec:index}
\begin{figure}
\centerline{\includegraphics[width=8cm]{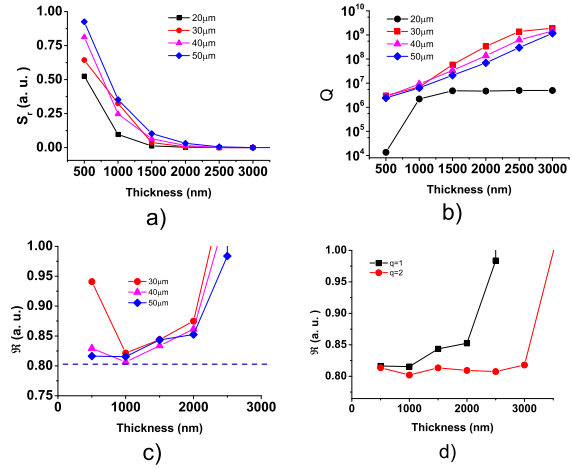}}
\caption{(a) Sensitivity is higher for thinner shells  while (b) Q-factors drop for microbubbles filled with water for  different bubble diameters (20 $\mu$m to 50 $\mu$m) and shell thicknesses (500 nm to 3 $\mu$m). For a certain thickness, an optimized resolution, defined by Eq. \ref{eq:res}, can be achieved in (c) (horizontal, blue dashed line shows the minimum). Diameters in the figure are: 20 $\mu$m (black squares), 30 $\mu$m (red circles), 40 $\mu$m (purple triangles) and 50 $\mu$m (blue spades). (d) Comparison of $\Re$ with $q=1$ (black squares) and $q=2$ (red circles) for the same microbubbles.}
\label{fig:index}
\end{figure}

 Observation of frequency shifting induced by subtle refractive index variations is the most common sensing mechanism for WGRs \cite{Arnold2003, Armani2007, Vollmer2008}, especially for LCORRs \cite{JonathanWard2013, White2006a, Suter2007, White2006, White2007, Li2010a}, where different materials can be injected into the interior core volume. A change in the core material due to changes in concentration, temperature, or pressure, causes a small change in the effective index of the mode. This is sufficient to generate a WGM frequency shift. Intuitively, for higher sensitivity, the light in the microbubble should be distributed in the inner core as much as possible. It can be qualitatively described by the following equation:
\begin{equation}
S=\frac{\partial \lambda}{\partial U}=\frac{\kappa_c}{n_c}\frac{\partial\lambda}{\partial U}+\frac{\kappa_s}{n_s}\frac{\partial\lambda}{\partial U}.
\label{eq:RIdefine}
\end{equation}
$\kappa_c$ and $\kappa_s$ are the proportion of EM field in the core and shell respectively while  $n_c$ and $n_s$ are their refractive indices. For simplicity, we assume that the physical quantity only changes the core material index, so that the sensitivity is proportional to $\kappa_c$ only. This means that the sensitivity follows the same trend for thickness as shown in Fig. \ref{fig:droplet}. To achieve high sensitivity, e. g. for the fundamental mode, it is better to have a very thin shell ($<$300 nm) where more than 80\% of the light is in the core; in other words, the bubble is in the quasi-droplet regime. However, in general, the core material has higher absorption (e. g. water) than the silica shell, thus limiting the Q-factor. Similar to Section \ref{sec:pressure}, an optimum resolution of the microbubble sensor exists, and it depends on the shell thickness. A precise value of sensitivity is obtained by FEM simulations. In order to implement numerical optimization that can deal with all kinds of index sensing, a detailed index-quantity relationship is not introduced. $S_r$, is defined by the ratio of frequency shift to refractive index change by $\partial\lambda/\partial n$. By introducing a small index change to the core material (0.001) in the COMSOL model, the frequency shift is calculated and plotted (Fig. \ref{fig:index}). To determine the Q-factor, it is assumed that the core is water with a high absorption at 1.55 $\mu$m  ($\epsilon_i=-3.577\times10^{-6}$ estimated from the water absorption coefficient $\alpha$ at 1.55 $\mu$m).\\
As discussed for Eq. \ref{eq:RIdefine}, the simulation shows the same result as in Fig. \ref{fig:index}(a), that $S_r$ increases with decreasing shell thickness. When the thickness is larger than 2.5 $\mu$m, most of the light is distributed in the shell, yielding a zero frequency shift. For different microbubble sizes, the sensitivity is slightly different. The Q-factors of the corresponding microbubbles are also calculated and presented in Fig. \ref{fig:index}(b). It should be mentioned that for a 20 $\mu$m diameter microbubble, the Q-factor is lower than for larger sizes. This is due to higher water absorption plus high radiation loss since light is not so well confined in the shell for such a small bubble. Therefore, in the following discussion, only diameters larger than 20 $\mu$m are considered. Using the results from Fig. \ref{fig:index}(a) and (b), optimized shell thicknesses and microbubble sizes for best resolution can be obtained. The optimized thickness is about 1 $\mu$m for microbubbles with diameters between 30-50 $\mu$m.  Around this thickness, the resolution does not vary too much with bubble size, which is also due to exponential dependencies of Q and sensitivity to shell thickness. The optimized microbubble for the fundamental first order mode is not within the quasi-droplet regime, due to a poor wavelength selection (1.55 $\mu$m). As has been discussed, the higher order modes are in the quasi-droplet regime over a larger range than the $q=1$ mode, so for thicknesses less than 2.5 $\mu$m the sensitivity will not drop too much. Using the same method, $\Re$ is calculated for the $q=2$ mode in a 50 $\mu$m diameter microbubble and plotted together with the $q=1$ mode in Fig. \ref{fig:index}(d). From this plot one can see that the resolution remains unchanged for thicknesses ranging from 500 nm to 3 $\mu$m. Therefore, in a practical sensing application, the second order mode is recommended since it decreases the difficulty of controlling the shell thickness when fabricating the device. The resolution can be further improved by using an alternative core material or a light source at a different wavelength where absorption loss is lower.\\
 So far it has been shown that high resolution refractive index sensing can be obtained by using a microbubble operating close to the quasi-droplet regime. It is assumed that index changes occur only in the core region. In some other situations, such as thermal sensing \cite{Suter2007}, changes in both shell and core refractive index must be considered. The thermo-optical coefficient of silica is positive, which leads to a red shift of the modes if the temperature rises. The core is often filled with a negative thermo-optical liquid, such as water, ethanol, or acetone. The net thermal shifting of a bubble is determined by the proportion of intensity in the shell and core, as in Eq. \ref{eq:RIdefine}. This can be tuned by selecting the shell thickness. Recent experimental results have proven that a thermally induced red shift of silica can be compensated for \cite{Suter2007} and it is even possible to obtain a large inverse blue shift \cite{JonathanWard2013}.\\
\subsection{Nanoparticle sensing}
\begin{figure}
\centerline{\includegraphics[width=8cm]{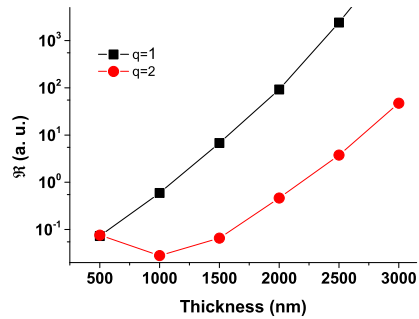}}
\caption{(a) Sensitivity of a microbubble for nanoparticle sensing. A relative frequency shift to the WGM is caused by a spherical nanoparticle (diameter 500 nm) attached to a water-filled 50 $\mu$m microbubble. Inset: schematic picture of the simulation condition. (b) $\Re$ versus shell thickness for a 50 $\mu$m microbubble for sensing the 500 nm nanoparticle. The axis of resolution is plotted on a log scale, which implies that the resolution improves nearly exponentially for a thinner shell for the first order fundamental mode (black squares). Here, the first order mode is plotted as black squares while the red circles represent the second order mode. Lines joining the data points are simply guides for the eye.}
\label{fig:particle}
\end{figure}

As an extension of the optimization method discussed in this paper, let us now consider nanoparticle sensing in microbubbles as a final example. Nanoparticle detection and bimolecular sensing have been realized in other WGRs \cite{Armani2007,Vollmer2008,Zhu2010} and have also been generally discussed in LCORRs \cite{Li2010}. Usually, the particles, the WGR, and the evanescent coupler are in an aqueous environment so that the particles can be delivered to the sensing devices. In practice, microbubbles can benefit from their hollow structure, so that various liquids can be passed inside the device while the optical readout occurs outside (by taper coupling, for example) without being influenced by the liquid. For a simple estimation of this effect, suppose a nanoparticle with a radius, $r_0$, is attached to the inner surface of the shell and the core is filled with water (see inset of Fig. \ref{fig:particle}(a)). The nanoparticle possesses a refractive index difference to water, $\Delta\epsilon(r_0)$ in permittivity. This small perturbation by the particle can cause a frequency shift, $\delta\omega$, as follows:
\begin{equation}
\frac{\Delta\omega}{\omega}=-\frac{\int d^3r\Delta\epsilon(|\mathbf{E}(\mathbf{r})|)^2}{2\int d^3r\epsilon(|\mathbf{E}(\mathbf{r})|)^2}
\label{eq:particle}
\end{equation}
On calculation, the frequency shift of the fundamental TE mode is calculated assuming that $\Delta\epsilon$ is a very small perturbation ($\Delta\epsilon=0.005$) in the above Eq. \ref{eq:particle}. The sensitivity is then plotted (Fig. \ref{fig:particle}(a)).\\
Since the nanoparticle changes the refractive index in the evanescent field that penetrates into the liquid core, for thinner shells the sensitivity is much higher. Subwavelength thickness is required for particle sensing. Otherwise the sensitivity goes to zero. The resolution determined from Eq. \ref{eq:res} and the Q-factor data presented in Fig. \ref{fig:index}(b) is plotted in Fig. \ref{fig:particle}(b). Resolution worsens exponentially when the shell thickness increases for a 50 $\mu$m microbubble. This is quite similar to the refractive index sensing situation, but with even more sensitive dependency on thickness. It can be understood by considering the concepts discussed in Section \ref{sec:droplet}. From Eq. \ref{eq:particle}, the nanoparticle is sensed by the value of $|\mathbf{E}(r)|^2$, which means that a high sensitivity is achieved when the radial maximum covers the position of the nanoparticle. As discussed in Section \ref{sec:droplet} and shown in Fig. \ref{fig:droplet}, if the microbubble is in the quasi-droplet regime, the maximum is shifted from inside the shell to the inner boundary of the microbubble near the position of the nanoparticle. If the relative position of the maximum to the nanoparticle changes, it leads to an exponential increase in $|\mathbf{E}(\mathbf(r))|^2$. This is the origin of exponentially improved resolution. Sensing nanoparticles with second order modes is also shown in Fig. \ref{fig:particle}. The simulation shows that there is an increase in sensitivity when the shell thickness is around 1.5 $\mu$m and a minimum in relative resolution at 1 $\mu$m. This corresponds to the multi maxima in the core discussed in Section \ref{sec:droplet}. It is also obvious that, for  microbubbles with the same shell thickness, both the sensitivity and resolution of the $q=2$ mode are better than for the $q=1$ mode. Within the simulation range from 500 nm to 1.5 $\mu$m, the microbubble is in the quasi-droplet regime for the $q=2$ mode while out of this regime for the $q=1$ mode, thereby proving that the quasi-droplet regime is quite important for high sensitivity particle sensing. Microbubbles in the quasi-droplet regime have other advantages. For example, for higher order modes in the quasi-droplet regime, more mode maxima lie in the core, implying a deeper penetration of the mode into the liquid. Even if one requires a method for sensing particles that are not attached to the inner surface, high sensitivity is still achievable if it is done using the appropriate higher order mode and a carefully designed shell thickness.  This is of more practical significance in biochemical sensing applications. \\
Here, the absolute frequency shift due to the presence of a single particle was not discussed since the specific material and geometrical properties of the nanoparticle were not assigned in our simulations. However, the calculation method used herein is universal and, therefore, it should be capable of simulating such a case. A complicated modification to introduce arbitrary nanoparticles near the surface of a toroidal cavity to break the axial symmetry has been reported \cite{Kaplan:13}. With some modifications, this method should also be suitable for microbubbles.\\
\section{Conclusions}
WGM optical properties of microbubble WGRs have been studied with numerical simulation results based on FEM. When the shell thickness diminishes to subwavelength scale, microbubbles operate in the so-called quasi-droplet regime, where WGMs are dominated by the presence of the liquid core. This provides an ultra-sensitive way to detect liquid optical properties.  Optimization was performed to achieve the best resolution for three types of sensing applications. This method can be further developed for a wide range of sensor optimization designs with microbubbles, such as newly developed optomechanical, microfluidic devices \cite{Bahl2013} and group velocity dispersion control for utilization in optical frequency comb generation \cite{Li:13}.\\
\bibliography{simulations}
\bibliographystyle{naturemag}
\begin{addendum}
 \item This work is supported by OIST Graduate University. We thank Dr. Yongping Zhang for fruitful discussions and Mr. Nitesh Dhasmana for his help in preparing this manuscript.
 \item[Correspondence] Correspondence and requests for materials
should be addressed to Y. Y. ~(email: yong.yang@oist.jp).
\end{addendum}

\end{document}